\documentclass[aps, preprint, floatfix, superscriptaddress, showpacs]{revtex4-1}

\usepackage{graphicx}
\usepackage{color}
\usepackage{calc}
\usepackage{array}
\usepackage{graphicx}
\usepackage{amsmath, amssymb}
\usepackage{natbib}
\usepackage{hyperref}

\def\abs#1{\left \vert #1 \right \vert}

\def\r{\bf r}

\DeclareMathOperator{\sech}{sech}

\begin{document}

\title{Effect of scattering lengths on the dynamics of a \\
       two-component Bose-Einstein condensate}

\author{G{\'{a}}bor \surname{Csire}}%
\affiliation{Institute of Physics,
             Budapest University of Technology and Economics, \\
             H-1111, Budafoki {\'{u}}t 8, Hungary}
\affiliation{Institute of Physics,
             E{\"{o}}tv{\"{o}}s Lor{\'{a}}nd University, \\
             H-1117, P{\'{a}}zm{\'{a}}ny P{\'{e}}ter s{\'{e}}t{\'{a}}ny 1/A, Hungary}

\author{D{\'a}niel Schumayer}
\affiliation{Jack Dodd Centre for Quantum Technology,
             Department of Physics, \\
             University of Otago,
             730 Cumberland St,
             Dunedin 9016,
             New Zealand}

\author{Barnab{\'a}s Apagyi}
\email{apagyi@phy.bme.hu} \affiliation{Institute of Physics,
             Budapest University of Technology and Economics, \\
             H-1111, Budafoki {\'{u}}t 8, Hungary}

\date{\today}

\begin{abstract}
   We examine the effect of the intra- and interspecies scattering
   lengths on the dynamics of a two-component Bose-Einstein condensate,
   particularly focusing on the existence and stability of solitonic
   excitations. For each type of possible soliton pairs stability ranges
   are presented in tabulated form. We also compare the numerically
   established stability of bright-bright, bright-dark and dark-dark
   solitons with our analytical prediction and with that of
   Painlev{\'e}-analysis of the dynamical equation. We demonstrate that
   tuning the inter-species scattering length away from the predicted
   value (keeping the intra-species coupling fixed) breaks the stability
   of the soliton pairs.
\end{abstract}


\maketitle

\section{Introduction \label{sec:introduction}}

Since Bose-Einstein condensates (BECs) can be routinely prepared in laboratories,
ultracold gases became a very important testbed for many predictions of condensed
matter physics \cite{Bloch2008}. The experimental examination of binary condensates
started nearly the same time as for single condensates by using two different
quantum states of the same species, such as $^{87}$Rb \cite{Myatt1997} or $^{23}$Na
\cite{Stamper-Kurn1998}. With the development of sympathetic cooling ultracold
mixtures have been assembled from two different alkalies, $^{41}$K--$^{87}$Rb
\cite{Modugno2001, Thalhammer2008}, $^{7}$Li--$^{133}$Cs \cite{Mudrich2002} and
$^{87}$Rb--$^{133}$Cs \cite{Pilch2009} or for different isotopes of the same alkali
atom $^{85}$Rb--$^{87}$Rb \cite{Papp2008}. The tunability of the inter- and intra-species
scattering lengths via driving the mixture through a Feshbach-resonance
has also been experimentally demonstrated \cite{Thalhammer2008, Pilch2009}.

The ability to create Bose-Einstein condensate(s), a highly coherent form of matter,
also facilitated the convergence of two fields of physics: condensed matter physics
and quantum optics, and therefore BECs became favourable candidates for examining
the effects of nonlinearity in matter waves, where this nonlinearity originates from
the mean-field representation of the interatomic interaction. The similarity between
electromagnetic waves in nonlinear medium and coherent matter waves is also expressed
in the equations of motion which is the nonlinear Schr{\"o}dinger equation (NLS) for
the former and the Gross-Pitaevskii (GP) equation for the latter. Although the physical
interpretation of these equations is different, their structures are the same apart
from the external potential term. Furthermore, in some cases this extra term can even
be removed~\cite{Schumayer2002} and the GP equation is transformed into a form
coinciding with the NLS equation. Consequently all results for the NLS equation
known in nonlinear optics can be readily adapted to Bose-Einstein condensates.

One of the surprising phenomena of nonlinear optics is the existence of particle-like
wave-forms, the so-called solitons~\cite{Hasegawa2002}. Such excitations have already
been experimentally observed in single- or two-component Bose-Einstein condensates:
dark solitons \cite{Burger1999, Denschlag2000}, bright solitons \cite{Khaykovich2002,
Strecker2002}, their two-component coupled analogues, the dark-dark \cite{Ohberg2001},
bright-bright \cite{Liu2002} or even dark-bright \cite{Busch2001, Becker2008}
multi-component solitary waves \cite{Ginsberg2005, Becker2008}.

However, the question of existence of solitons needs more attention than simply
recognising the similarity between the two governing equations. The existence of
solitons is strongly related to the integrability of the given physical model.
An usual test to determine whether or not an equation is integrable is the
Painlev{\'e}-test (P-test) \cite{Steeb1984}. It was shown that without any potential
term or inhomogeneity the one-dimensional NLS equation, $i u_{t} + u_{xx} \pm 2
\abs{u}^{2} u = 0$, is completely integrable both in its one-component
\cite{Zakharov1972, Steeb1984} or multi-component \cite{Zakharov1974, Zakharov1985}
form. However, the inclusion of a potential term, $v(x,t) u$, in the one-component
NLS equation or different coupling strengths in the multi-component NLS/GP equations
fundamentally changes their integrability \cite{Sahadevan1986, Clarkson1988}. It
was shown that integrability is preserved provided the external potential, $v(x,t)$,
has a specific form \cite{Clarkson1988}. The authors examined \cite{Schumayer2001}
the integrability of the two-component coupled Gross-Pitaevskii (CGP) equations and
lead to similar conclusion: the scattering lengths and the external potentials cannot
be arbitrary if the integrability of CGP is to be preserved. The inter- and intraspecies
scattering lengths must satisfy the following equation
\begin{equation} \label{eq:ConditionForIntegrability}
   \frac{%
         2 \xi_{1} \xi_{2}
         - \kappa_{1} \xi_{1}
         - \kappa_{2} \xi_{2}
        }%
        {
         \xi_{1} \xi_{2}
         - \kappa_{1} \kappa_{2}
        }
   =
   \!\frac{(2n+1)^{2} + 7}{16},
\end{equation}
where $\xi_{1}=a_{11}/a_{21}$, $\xi_{2}=a_{22}/a_{12}$, $\kappa_{1}=\mu_{11}/ \mu_{21}$,
$\kappa_{2}=\mu_{22}/ \mu_{12}$ and $\mu_{ij}$ denotes the reduced mass of pair of
particles composed by an atom from the $i$th and $j$th species. On the right hand side
of Eq.~(\ref{eq:ConditionForIntegrability}) $n$ is a non-negative integer number. One
may call $n$ a {\emph{classification number}}, because it determines the form of the
external potentials for which CGP equations remain integrable. For example, for $n=2$,
the external potential, apart from the quadratic trapping potential, may even contain
an imaginary time-dependent term \cite{Schumayer2001}. This term can mimic the loss or
gain in the number of particles of the given species. We note here that usually
dissipation works against long-living coherent matter waves, however, the importance of
this imaginary potential term has been analysed in \cite{Brazhnyi2009, Rajendran2009}
and shown to permit exact soliton solution \cite{Nakkeeran2001}. In the context of BECs
at finite temperature, the interaction of the condensate with the thermal cloud could
also be taken into account as an imaginary term in the governing GP equation. This
interaction, due to its stochastic nature, can influence the dynamics of the solitons,
via density-fluctuation.

In this paper we carry out an analysis on how the intra- and interspecies interaction
influences the dynamics of a binary mixture of Bose-Einstein condensates.  We select
out of the many possible systems of two component BECs the pairs $^{87}$Rb--$^{87}$Rb
(prepared in two distinct hyperfine states), $^{23}$Na--$^{87}$Rb, and $^7$Li--$^{39}$K.
In the former two systems there is a possibility to study stability of bright-dark and
dark-dark soliton pairs, while the last pair is capable to sustain bright-bright
and bright-dark excitations.

The organization of the paper is as follows. Section~\ref{sec:TheModel} defines a quasi
one-dimensional model derived from the general three-dimensional, coupled Gross-Pitaevskii
equations assuming cigar-like harmonic oscillator trap potential. In the first part of
section \ref{sec:StabilityTest} we shall perform a stability analysis based on coupled
soliton excitations, and on Eq.~(\ref{eq:ConditionForIntegrability}) of the P-test. In
the second part of section \ref{sec:StabilityTest}, possible new interesting modes
exhibited by the bright-bright solitons will be shown. Section \ref{sec:summary} is
devoted to a conclusion and the summary.

\section{The model \label{sec:TheModel}}

In ultracold gases the interaction between two particles can usually be well
described by a scalar parameter, the scattering length. For a two-component
Bose-Einstein condensate one has to introduce three, possibly different,
scattering lengths, characterising the intra-species interactions ($a_{11}$,
$a_{22}$) and the inter-species coupling ($a_{12} = a_{21}$).

In the mean-field approximation a two-component BEC is described by the
coupled Gross-Pitaevskii equations \cite{Gross1961, Pitaevskii1961} in
the form:
\begin{equation} \label{eq:coupled_GP_math_notation}
   i \hbar \, \frac{\partial}{\partial t} \Psi_{i}
   =
   \left \lbrack
      - \frac{\hbar^{2}}{2m_{i}} \, \Delta
      + \sum_{j=1}^{2}{\Omega_{ij} \abs{\Psi_{j}}^{2}}
      + V_{i}
   \right \rbrack \! \Psi_{i},
\end{equation}
where $m_{i}$ denotes the individual mass of the $i$th atomic species, $\Omega_{ij}
= 2 \pi \hbar^{2} a_{ij} /\mu_{ij}$ with $a_{ij}$ being the 3D scattering length,
$\mu_{ij}=m_{i} m_{j}/(m_{i} + m_{j})$ is the reduced mass, and $V_{i}$ denotes
the external trapping potential. In the following, indices $i$ and $j$ label the
components, therefore, take only two values, 1 and 2. In the case of real trap
potentials the normalisation of the wave functions reads as $N_{i} = \int{\!\!
\abs{\Psi_{i}}^{2} \! dV}$ with $N_{i}$ denoting the number of atoms in the $i$th
component. We exclude those cases from our analysis where the species can transform
into each other, therefore the number of atoms in each component hereafter is
conserved.

\subsection{Transformed equations \label{subsec:TransformedEquations}}

If the three dimensional quadratic trapping potential is weak in one direction, i.e.
\begin{equation} \label{eq:3D-HO}
   V_{i}
   =
   \frac{1}{2} m_{i}
   \Bigl \lbrack
      \omega_{i, x}^{2}      x^{2} +
      \omega_{i, \perp}^{2} \left ( y^{2} + z^{2} \right )
   \Bigr \rbrack,
\end{equation}
where $\omega_{ix} \ll \omega_{i\bot}$, one may replace the three-dimensional equations
(\ref{eq:coupled_GP_math_notation}) with a coupled system of quasi one-dimensional GP
equations. Although Eq.~(\ref{eq:coupled_GP_math_notation}) is a nonlinear equation,
physically we may assume that the weak $x$-direction decouples from the strong [$yz$]--plane,
therefore the macroscopic wave functions can be written as
\begin{equation} \label{eq:1D-wf}
   \Psi_{i} ({\r},t)
   =
   \sqrt{N_{1}}\,
   \psi_{i} (x,t)\, \chi_{i, \perp} (y,z,t)
\end{equation}
 with
$\chi_{i, \perp}$ represents the ground-state solution of the corresponding two-dimensional
Schr{\"o}dinger equation in the [$yz$]--plane. The external potential introduces suitable
units of length and time as $a_{\perp} = \sqrt{ {\hbar}/{m_{1} \omega_{1, \perp}}}$ and
$\tau = 1 / \omega_{1, \perp}$, respectively. By rescaling the spatial and temporal variable
with $a_{\perp}$ and $\tau$ one obtains two quasi one-dimensional GP equations
\begin{subequations}
   \begin{eqnarray}
      \label{eq:coupled_GP1}
      i {\psi}_{1, t}
      &=&
      \!\!
      \left \lbrack
         - \frac{1}{2} \, \partial_{xx}
         +  \frac{\lambda_1^2}{2} \,  x^2
         + {b_{11} \abs{{\psi}_{1}}^{2}}
         + {b_{12} \abs{{\psi}_{2}}^{2}}
      \right \rbrack \! {\psi}_{1},
      \hspace*{7mm}
      \\
   \label{eq:coupled_GP2}
      i {\psi}_{2,t}
      &=&
      \!\!
      \left \lbrack
         - \frac{\kappa}{2} \, \partial_{xx}
         +  \frac{\lambda_2^2}{2\kappa} \,  x^2
         + {b_{21} \abs{{\psi}_{1}}^{2}}
         + {b_{22} \abs{{\psi}_{2}}^{2}}
      \right \rbrack \! {\psi}_{2},
   \end{eqnarray}
\end{subequations}
where $b_{11} = 2a_{11} N_{1}$, $b_{22} = 2 a_{22} N_{1} \kappa/
\gamma$, $b_{12} = b_{21} = 2 a_{12} N_{1} (1 + \kappa)/(1 +
\gamma)$, $\gamma = \omega_{2, \perp}/ \omega_{1, \perp}$, $\kappa
= m_{1}/m_{2}$, $\lambda_{1} = \omega_{1,x} / \omega_{1, \perp}$,
$\lambda_{2} = \omega_{2,x} / \omega_{1, \perp}$. The
normalisation is such that $\int{\abs{\psi_{1}}^{2}dx} = 1$ and
$\int{\abs{\psi_{2}}^{2}dx} = N_{2}/N_{1}$. Moreover, the relation
$\gamma^{2} = \kappa$ must hold if both species experience the
same harmonic potential. (Note the slight departure from
Ref.~\cite{Liu2009} in the definition of $b_{22}$ which, however,
may result in large difference of values of $b_{22}$ if a
two-component condensate contains species with different masses
$m_{1} \ne m_{2}$.)

\subsection{Thomas-Fermi background \label{sec:TFBackground}}

If the kinetic energy term is negligible compared to the potential
energy terms in (\ref{eq:coupled_GP1}-b), then one may apply the
Thomas-Fermi approximation to determine the density distribution
of the ground state. Following~\cite{Schumayer2004} we write the
corresponding wave-functions as
\begin{equation} \label{eq:TF-1}
   \psi_i(x,t)
   \approx
   \Phi^{\mathrm{TF}}_{i} (x)
   \exp{(-i E_{i}^{\mathrm{TF}} t/ \hbar)},
\end{equation}
resulting in TF densities
\begin{equation} \label{eq:TF-2}
   \abs{\Phi^{\mathrm{TF}}_{i}}^{2}
   =
   \frac{A_{i}}{\Delta}
   \left ( x_{i}^{2} - x^{2} \right ),
   \quad
   (\abs{x} < x_{i})
\end{equation}
and TF energies
\begin{subequations}
   \begin{eqnarray}
      \label{eq:TF-E1}
      E_{1}^{\mathrm{TF}}
      &=&
      \left ( b_{11} A_{1} x_{1}^{2} + b_{12} A_{2} x_{2}^{2} \right ) / \Delta,
      \\
      \label{eq:TF-E2}
      E_{2}^{\mathrm{TF}}
      &=&
      \left ( b_{12} A_{1} x_{1}^{2} + b_{22} A_{2} x_{2}^{2} \right ) / \Delta,
   \end{eqnarray}
\end{subequations}
where $\Delta = b_{11} b_{22}-b_{12}^{2}$. The parameters $A_i$ and $x_{i}$ represent
the amplitude of the density and the extension of the condensates, respectively. All
these quantities are determined by the system parameters $b_{ij}$, $N_{i}$, $\lambda_{i}$
according to the following relations:
\begin{equation} \label{eq:A-1}
   A_{1} = \frac{b_{22}}{2} \lambda_{1}^{2} - \frac{b_{12}}{2\kappa} \lambda_{2}^{2},
   \quad \mbox{and} \quad
   A_{2} = \frac{b_{11}}{2 \kappa} \lambda_{2}^{2} - \frac{b_{12}}{2} \lambda_1^2,
\end{equation}
while the extensions are
\begin{equation} \label{eq:x-1}
   x_{1} = \left ( \frac{3}{4} \frac{\Delta}{A_{1}} \right )^{1/3}
   \quad \mbox{and} \quad
   x_{2} = \left ( \frac{3}{4} \frac{\Delta}{A_{2}} \frac{N_{2}}{N_{1}} \right )^{1/3}.
\end{equation}
Although the Thomas-Fermi density distribution is not physical at $x=x_{i}$,
it still provides a good starting point for analytical calculations. In our numerical
treatment we will not use this approximation, rather start our simulations from the
appropriate ground state solution of the one-dimensional GP equation.

\subsection{Coupled soliton excitations \label{sec:CoupledSolitons}}

Now we are seeking solutions of Eqs.~(\ref{eq:coupled_GP1}-b) which support
soliton excitations. A static soliton excitation can be written as
\cite{Schumayer2004}
\begin{equation} \label{sol-excit}
   {\widetilde{\psi}}_{i} (x,t)
   =
   \Phi^{\mathrm{TF}}_i(0)
   \varphi_{i} (x)
   \exp{(-i {\widetilde{E}}_{i} t)}.
\end{equation}
By inserting this ansatz into the GP Eqs.~(\ref{eq:coupled_GP1}-b) and
neglecting the small potential contributions one obtains the coupled soliton
equations as follows
\begin{subequations}
   \begin{eqnarray}
      \label{eq:coupled_SOL1}
      \widetilde{E}_{1} {\varphi}_{1}
      &=&
      \left \lbrack
         - \frac{1}{2} \, \partial_{xx}
         + \widetilde{b}_{11} \abs{{\varphi}_{1}}^{2}
         + \widetilde{b}_{12} \abs{{\varphi}_{2}}^{2}
      \right \rbrack \! {\varphi}_{1}
      \\
      \label{eq:coupled_SOL2}
      \widetilde{E}_2 {\varphi}_{2}
      &=&
      \left \lbrack
       - \frac{\kappa}{2} \, \partial_{xx}
       + \widetilde{b}_{21} \abs{{\varphi}_{1}}^{2}
       + \widetilde{b}_{22} \abs{{\varphi}_{2}}^{2}
      \right \rbrack \! {\varphi}_{2}
   \end{eqnarray}
\end{subequations}
with $\widetilde{b}_{ij} = b_{ij} A_{j} x_{j}^{2} / \Delta$. The normalisation
of the soliton solutions reads as follows
\begin{subequations}
   \begin{eqnarray}
      \label{eq:norm_SOL1}
      \int_{-L_{1}}^{L_{1}}{\abs{\varphi_{1}}^{2} dx}
      &=&
      \frac{\Delta}{A_{1} x_{1}^{2}},
      \\
      \label{eq:norm_SOL2}
      \int_{-L_{2}}^{L_{2}}{\abs{\varphi_{2}}^{2} dx}
      &=&
      \frac{\Delta}{A_{2} x_{2}^{2}}
      \frac{N_{2}}{N_{1}},
   \end{eqnarray}
\end{subequations}
where the integrations, in both cases, are over the spatial extension,
$L_{i}$, of the solitons.

The above coupled equations admit generic moving soliton solutions
of the types: bright-bright (BB), bright-dark (BD), dark-bright (DB)
and dark-dark (DD). We shall investigate here a simple static bright-dark
soliton pair solution by taking the first component to be a static bright
soliton
\begin{subequations}
   \begin{equation} \label{eq:_BSOL1}
      \varphi_{1}^{\mathrm{BD}} (x) = q_{1} \sech{(k_{1} x)},
      \quad
      \varphi_{1}^{\mathrm{BD}}(x \rightarrow \pm \infty) = 0,
      \quad
   \end{equation}
and the second component to be the static dark soliton
   \begin{equation} \label{eq:_BSOL2}
      \varphi_{2}^{\mathrm{BD}} (x) = q_{2} \tanh{(k_{2} x)},
      \quad
      \varphi_{2}^{\mathrm{BD}}(x \rightarrow \pm \infty) = \pm q_{2}
   \end{equation}
\end{subequations}
with yet unknown complex amplitudes $q_{i}$ and wave-vector, $k_{i}$. The
latter one is related to the width of the soliton $k_{i} \sim 1/L_{i}$.
By inserting the above ansatz into the Eqs.~(\ref{eq:coupled_SOL1}-b)
and equating the coefficients of the constant and $x$-dependent terms,
respectively, one may conclude that the wave-vectors of the dark and
bright solitons must be equal, $k_{1} = k_{2} \equiv k$. The amplitudes
are expressed by the system parameters as
\begin{subequations}
   \begin{eqnarray}
      \label{eq:_q1}
      \abs{q_{1}}^{2} = \frac{k^{2}}{A_{1} x_{1}^{2}} (\kappa b_{12} - b_{22}),
      \\
      \label{eq:_q2}
      \abs{q_{2}}^{2} = \frac{k^{2}}{A_{2} x_{2}^{2}} (\kappa b_{11} - b_{12}).
   \end{eqnarray}
\end{subequations}
The energy of these excitations read as
\begin{subequations}
   \begin{eqnarray}
      \label{eq:_tildeE1BD}
      \widetilde{E}_{1}^{\mathrm{BD}}
      &=&
      k^{2} \left (
               \frac{\kappa b_{11} - b_{12}}{\Delta} b_{12} - \frac{1}{2}
            \right ),
      \\
      \label{eq:_tildeE2BD}
      \widetilde{E}_{2}^{\mathrm{BD}}
      &=&
      k^{2} \, \frac{\kappa b_{11} - b_{12}}{\Delta}b_{22}.
   \end{eqnarray}
\end{subequations}

Suppose now that the width parameter $k$ is a real number. The
modulus of the amplitudes of the bright-dark soliton pair superimposed
on the Thomas-Fermi background must be real numbers, thus one obtains
a the following set of conditions for the existence of this bright-dark
soliton excitations
\begin{eqnarray}
      \label{eq:_BD1cond}
      C_{1} \equiv \frac{\kappa b_{12} - b_{22}}{\Delta} \ge 0
      \quad \mbox{and} \quad
      C_{2} \equiv \frac{\kappa b_{11} - b_{12}}{\Delta} \ge 0.
      \qquad
   \end{eqnarray}

One may apply the same method to generate static bright-bright or
dark-dark soliton excitations. In the bright-bright case one obtains
the following solutions
\begin{subequations}
   \begin{eqnarray}
      \label{eq:_bb1}
      \varphi_{1}^{\mathrm{BB}} (x)
      &=&
      \frac{k}{\sqrt{A_{1}} x_{1}} \sqrt{\kappa b_{12} - b_{22}}\, \sech{(kx)},
      \\
      \label{eq:_bb2}
      \varphi_{2}^{\mathrm{BB}} (x)
      &=&
      \frac{k}{\sqrt{A_{2}} x_{2}} \sqrt{b_{12} - \kappa b_{11}}\, \sech{(kx)},
   \end{eqnarray}
\end{subequations}
with the conditions
\begin{equation}
   \label{eq:_BBcond}
   C_1\ge 0 \quad \textrm{and} \quad C_2\le 0,
\end{equation}
while the energies are $\widetilde{E}_{1}^{\mathrm{BB}} =
\widetilde{E}_{2}^{\mathrm{BB}}/\kappa  = -k^{2}/2$. The
dark-dark coupled soliton solutions read as follows
\begin{subequations}
   \begin{eqnarray}
      \varphi_{1}^{\mathrm{DD}} (x)
      &=&
      \frac{k}{\sqrt{A_{1}} \, x_{1}} \sqrt{b_{22} - \kappa b_{12}} \, \tanh{(kx)},
      \\
      \varphi_{2}^{\mathrm{DD}} (x)
      &=&
      \frac{k}{\sqrt{A_{2}} \, x_{2}} \sqrt{\kappa b_{11} - b_{12}} \, \tanh{(kx)},
   \end{eqnarray}
\end{subequations}
with the conditions
\begin{equation}
   \label{eq:_DDcond}
   C_{1} \le 0
   \quad \mathrm{and} \quad
   C_{2} \ge 0,
\end{equation}
and energies $\widetilde{E}_{1}^{\mathrm{DD}} =
\widetilde{E}_{2}^{\mathrm{DD}}/\kappa = k^{2}$.
Our interesting result shows that in the bright-bright and dark-dark
cases the energies are uniquely determined by the wave-vector and the
mass ratio. Note that the existence conditions (\ref{eq:_BD1cond}),
(\ref{eq:_BBcond}), and (\ref{eq:_DDcond}) are just the same as obtained
in Ref. \cite{Liu2009} for the existence of moving soliton pairs, while
the constraints for static excitations were published in \cite{Schumayer2004}.

\section{Stability tests by simulation \label{sec:StabilityTest}}

Below we are going to numerically investigate the stability of soliton
pairs. To solve the time dependent coupled Gross-Pitaevskii equations
(\ref{eq:coupled_GP1}-b), a third-order accurate split-step Fourier
transform method is used as described in Ref. \cite{Javanainen2006} for
a single-component condensate. Here we  solve the time independent
coupled Gross-Pitaevskii equations for their numerically exact ground
states using imaginary time method \cite{Lehtovaara2007} combined with
the split-step operator technique. Choosing initial distributions is
necessary to this method, and the Thomas-Fermi approximate solution
proved to be an effective initial guess for this purpose.

The procedure explained in the previous section can be generalised
for solitons moving with velocity $v$. Such a solution is given by
\begin{widetext}
\begin{subequations}
   \begin{eqnarray}
      \widetilde{\psi}_{1}^{\mathrm{BD}}
      &=&
      \Phi^{\mathrm{TF}}_1 (0)
      \varphi_{1}^{\mathrm{BD}} (x-vt)\,
      \exp{\left(-i {\widetilde{E}}_{1}^{\mathrm{BD}} t\right)}
      \exp{\!\left \{
                    -i \left \lbrack
                          v^{2} t \left ( \frac{b_{12} C_{2}}{\kappa^{2}} - \frac{1}{2} \right ) -
                          v (x - vt)
                       \right \rbrack
          \right \},
          }
      \hspace*{5mm}
      \\
      \widetilde{\psi}_{2}^{\mathrm{BD}}
      &=&
      \left \lbrack
         i \frac{\sqrt{C_2}}{\kappa}\,v + \Phi^{\mathrm{TF}}_2 (0) \varphi_{2}^{\mathrm{BD}} (x-vt)
      \right \rbrack \,
      \exp{\left(-i {\widetilde{E}}_{2}^{\mathrm{BD}} t\right)}
      \exp{\!\left (
                    -i \,\frac{b_{22} C_{2}}{\kappa^{2}} v^{2} t
          \right )
          }.
   \end{eqnarray}
\end{subequations}
\end{widetext}
At $v=0$ we obtain  the static soliton excitation solution as given by
\eqref{sol-excit} in the preceding section.

The existence conditions (\ref{eq:_BD1cond}) prescribe various relations
between domains of the inter- and intra coupling strengths, $b_{ij}$, where
it is possible to create bright-dark soliton pairs. These domains are listed
in Table \ref{tab:ExistenceOfBDSolitons}. Although the creation of a bright
soliton is generally associated with attractive interaction ($b_{ii} < 0$)
among the particles, Table \ref{tab:ExistenceOfBDSolitons} clearly shows
that, due to the appropriate other couplings, it may also be possible to
create a bright soliton in case of repulsive interaction $(b_{ii}>0)$. Such
a situation occurs in case 4a of Tab.~\ref{tab:ExistenceOfBDSolitons} which
we are going to analyse below.

\subsection{Moving bright-dark soliton pairs \label{sec:MovingBDSolitonPairs} }

Let us now investigate the stability of a bright-dark soliton pair for
the case 4a of Table \ref{tab:ExistenceOfBDSolitons} by considering two
experimentally accessible two-component BECs. The first is composed of
the two hyper-fine states of $^{87}$Rb atoms, the second is obtained
from $^{23}$Na and $^{87}$Rb atoms. In both cases the scattering lengths
are well known and can be tuned over a broad limit by using the Feshbach
resonance method. Our aim here is to explore the sensitivity of the temporal
evolution of soliton pair when the inter-atomic coupling strength, $b_{12}$,
is varied around the value prescribed by the ratio formula
(\ref{eq:ConditionForIntegrability}) obtained by performing
a Painlev{\'e}-analysis of the coupled GP equations \cite{Schumayer2001}.
In this respect we fix the intra-atomic strengths, $b_{ii}$, to values
easily accessible to the experiments and vary the inter-atomic interaction
values, $b_{12}$, within a small range allowed by the case 4a listed in
Table~\ref{tab:ExistenceOfBDSolitons}. In order to represent a more realistic
situation, a small velocity of $v=0.04$ is given the solitons, a weak harmonic
trapping potential is added along the longitudinal direction and the  solitons
are superimposed on the ground state density distribution. This procedure
spatially confines the solitons without affecting their essential stability
properties \cite{Liu2009}.

\begin{table}[h]
   \begin{tabular}{r<{\hspace*{1mm}}|p{5mm}p{5mm}cc}
      Case & $b_{11}$ & $b_{22}$ & Constraint on $b_{12}$ \\
      \hline\hline
       1   & $+$ & $-$ & no bright-dark soliton pair\\
       2   & $-$ & $+$ & $b_{11}\kappa<b_{12}<b_{22}/\kappa$\\
       3a  & $-$ & $-$ & $-\sqrt{b_{11}b_{22}}<b_{12}<b_{11}\kappa$ & if $\kappa^2 b_{11}>b_{22} $\\
       3b  &      &      & $b_{11}\kappa<b_{12}<-\sqrt{b_{11}b_{22}}$ & if $\kappa^2 b_{11}<b_{22} $\\
       4a  & $+$ & $+$ & $\sqrt{b_{11}b_{22}}<b_{12}<b_{22}/\kappa$ & if $\kappa^2 b_{11}<b_{22} $\\
       4b  &      &      & $b_{22}/\kappa<b_{12}<\sqrt{b_{11}b_{22}}$ & if $\kappa^2 b_{11}>b_{22} $\\
   \end{tabular}
   \caption{\label{tab:ExistenceOfBDSolitons}%
            Various domains of inter- and intra-atomic interaction strengths,
            $b_{ij}$, permitting the existence of a bright-dark soliton pair.}
\end{table}

\begin{figure}[hbt!]
   \includegraphics[width=\textwidth/2-3mm]{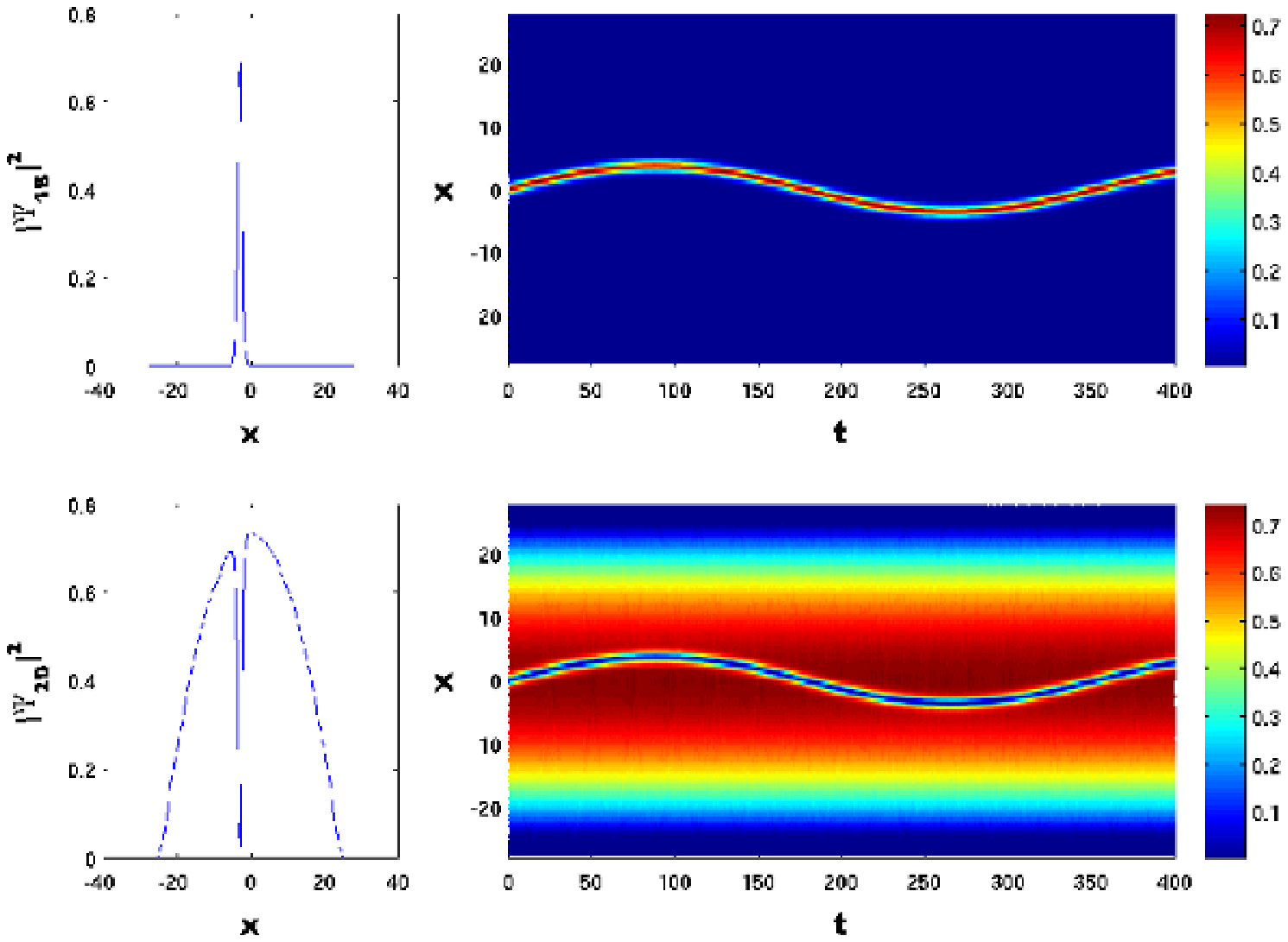}
   \includegraphics[width=\textwidth/2-3mm]{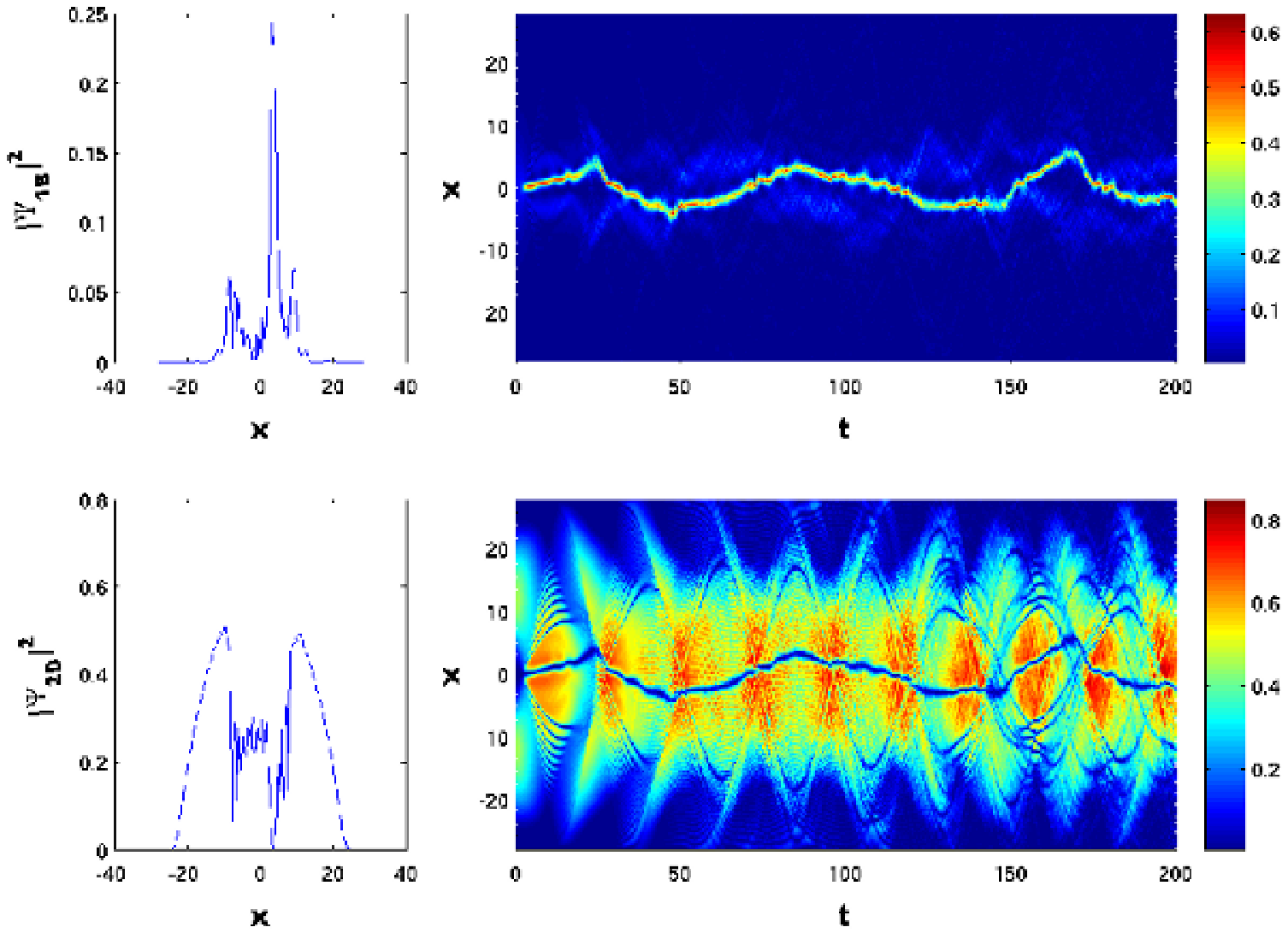}
   \caption{\label{fig:RbRb}%
            (Color online) Evolution of the bright-dark solitons of two-component
            BEC composed of two hyperfine states of the $^{87}$Rb atom. Intraspecies
            scattering length $a_{12}=5.5$\,nm ($b_{12}=13.6$) satisfying ratio
            formula (\ref{eq:ConditionForIntegrability}) (top two panels), $a_{12}=5.4$\,nm
            ($b_{12}=13.35$) disobeying the ratio formula (\ref{eq:ConditionForIntegrability})
            (bottom two panels). Other parameters $a_{11}=0.8 \times 5.5$\,nm ($b_{11}=10.88$),
            $a_{22}=1.2 \times 5.5$\,nm ($b_{22}=16.32$), $m_{1} = m_{2} = 87$\,au,
            $\omega_{1, \perp}=2\pi\times 710$ Hz, $\lambda_{1} = \lambda_{2}=0.2$,
            $v = 0.04$, $N_{1} = 500$, $N_{2}=6600$. The snapshots are depicting the
            soliton pairs at $t=20$ and $t=200$, respectively.}
\end{figure}

\begin{figure}[hbt!]
   \includegraphics[width=\textwidth/2-4mm]{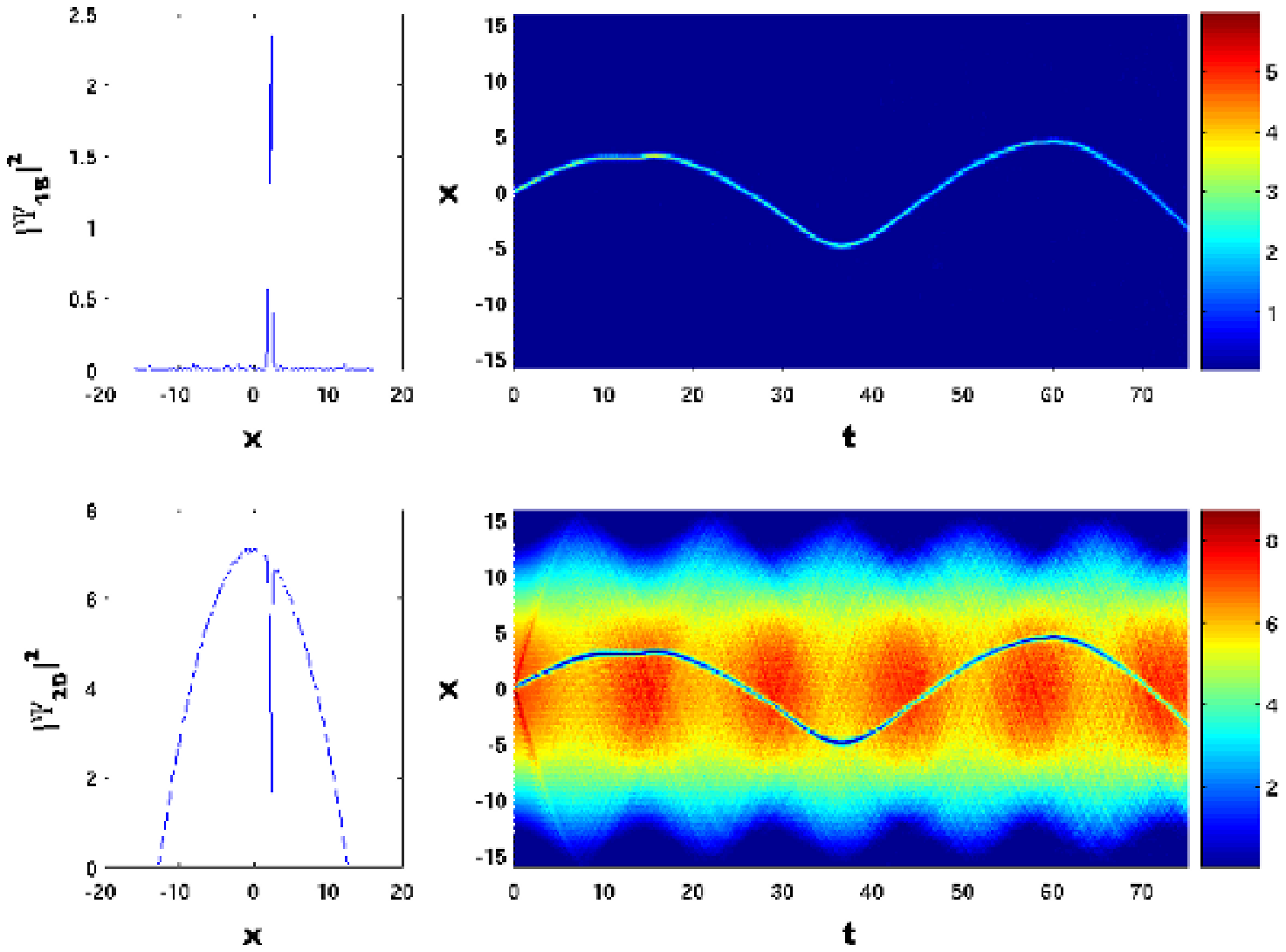}
   \includegraphics[width=\textwidth/2-4mm]{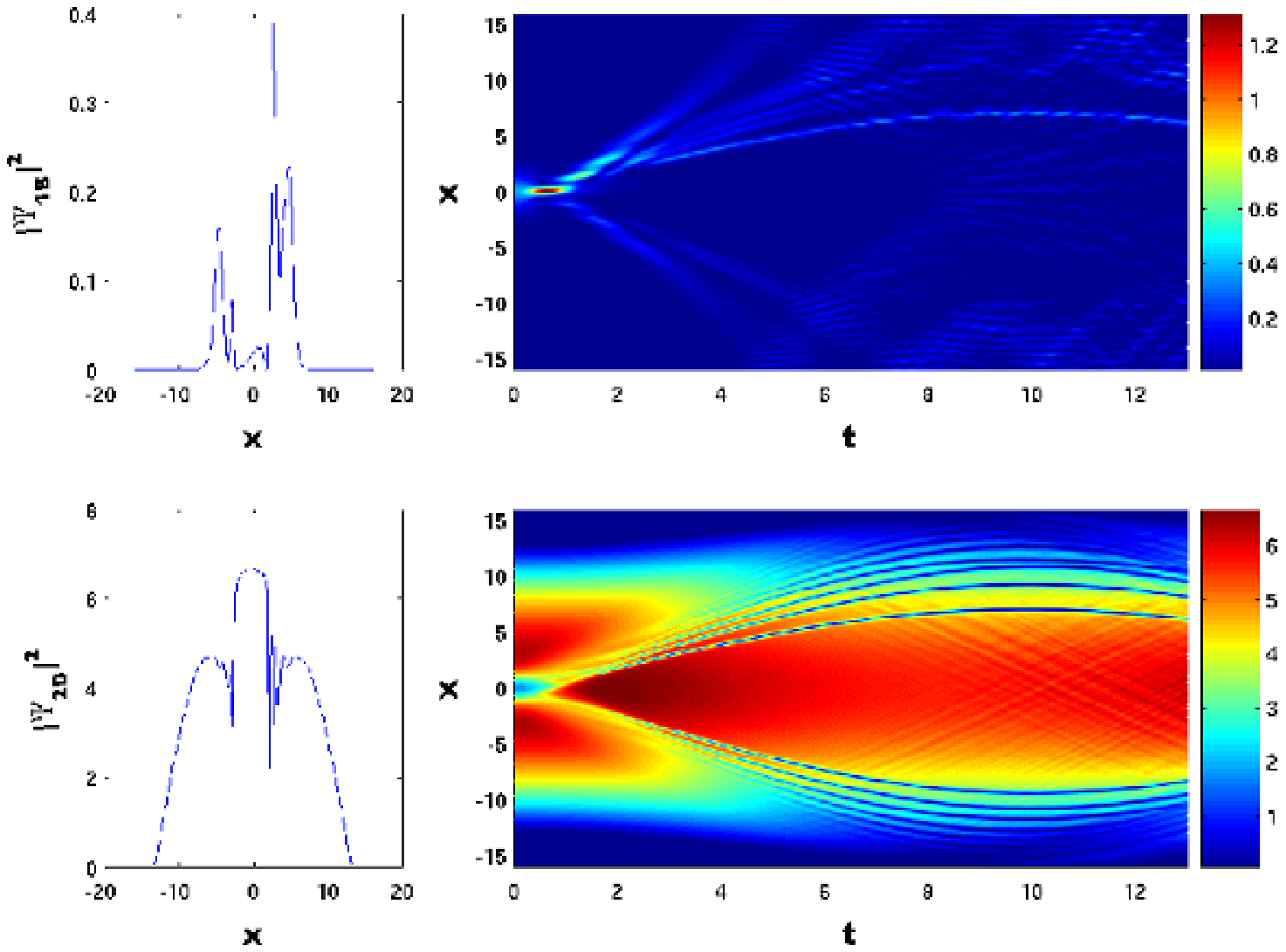}
   \caption{\label{fig:NaRb}%
            (Color online) Evolution of a bright-dark soliton pair in a two-component
            BEC composed of $^{23}$Na and $^{87}$Rb atoms. Intraspecies scattering
            length $a_{12}= 6.831$\,nm ($b_{12}=7.25$) satisfies formula
            (\ref{eq:ConditionForIntegrability}) (top two panels), while
            $a_{12}= 3.769$\,nm ($b_{12}=4$) deviates from Eq.~(\ref{eq:ConditionForIntegrability})
            (bottom two panels). Other parameters are: $a_{11}=2.7$\,nm $(b_{11}=3.4)$,
            $a_{22}= 5.5$\,nm $(b_{22}=3.6)$, $m_{1}=23$\,au, $m_{2}=87$\,au,
            $\omega_{1, \perp}=2 \pi \times 710$\,Hz, $\lambda_{1} = \lambda_{2}=0.25$,
            $v=0.15$, $N_{1}=500$, $N_{2}=58000$. The snapshots are taken at $t=20$,
            and $t=2$, respectively.}
\end{figure}

In Fig.~\ref{fig:RbRb}, we plotted a bright-dark soliton pair composed of
two hyperfine states of $^{87}$Rb atoms. The top two panels show the case
when the inter-species interaction is chosen according to
Eq.~(\ref{eq:ConditionForIntegrability}). Both solitons oscillate in the
harmonic trap in a stable manner with an angular frequency slightly less
than $\omega_{x}/\sqrt{2}$ what the one-dimensional Thomas-Fermi model
predicts for a single dark soliton \cite{Busch2000}. The difference is
probably caused by the presence of the bright soliton, since the bright
component fills the dip of the dark soliton, therefore the dark soliton
has to drag this extra mass as well. This effect has recently been observed
\cite{Becker2008} with a $^{87}$Rb-$^{87}$Rb condensate, prepared in the
$\vert F=2, m_{F} = 0 \rangle$ (bright soliton) and  $\vert F=1, m_{F} = 0
\rangle$ (dark soliton) hyperfine states.

However, if $a_{12}$ (or equivalently $b_{12}$) is tuned away from this
particular value, the stability is lost, and the initial forms of the
solitons are destroyed by the destructive interference of the constantly
emitted and re-captured sound waves. It is worthwhile to mention, although
only as a qualitative statement, that the appearance of sound waves made
the solitons' oscillation faster (see the different range of time in the
top two and bottom two panels of Fig.~\ref{fig:RbRb}). The sound waves
travel faster than the solitons, and after being reflected back from the
edge of the condensate, they collide with the solitons. The subsequent
collisions possibly speed up the oscillation and turn it into an irregular
sloshing. Despite the irregular movement of the solitons, the dark component
still captures the bright soliton during the motion.

In Fig.~\ref{fig:NaRb} the evolution of a bright-dark soliton pair in binary
BEC composed of $^{23}$Na and $^{87}$Rb atoms is plotted with corresponding
snapshots of the density distribution. Upper two panels exhibit the bright
($^{23}$Na) and dark ($^{87}$Rb) excitations, respectively, which are stable
for a long period due to the fine tuning of scattering lengths $a_{ij}\,
(i,j=1,2)$ which satisfy the P-test formula \eqref{eq:ConditionForIntegrability}.
On the lower two panels of Fig.~\ref{fig:NaRb} we see, however, that the
initial soliton excitations do not remain stable but are rapidly dissolved
due to detuning the interspecies scattering length $a_{12}$ from the value
obeying the integrability condition expressed by Eq. \eqref{eq:ConditionForIntegrability}.

Summarising, both examples exhibited in Figs.~\ref{fig:RbRb} and \ref{fig:NaRb}
show that a long lived stability of the bright-dark soliton pairs can be achieved
only in case if the interatomic interaction parameters $a_{12}$ (or $b_{12}$)
is chosen according to Eq.~\eqref{eq:ConditionForIntegrability}. This observation
emphasises that there may be situations in the creation of two-component BECs
when fine tuning of scattering lengths according to the P-test formula
(\ref{eq:ConditionForIntegrability}) may prove useful.

At the end of this section we briefly mention an early investigation of the
stability of a heterogeneous two-component Bose-Einstein condensate by Law
et al. \cite{Law1997}. This work carried out a linear stability analysis by
calculating the lowest eigenvalue of an excitation. In this approach the
appearance of a negative eigenvalue signals instability. It was shown for
a sodium-rubidium system, just as above, that stability occurs only in finite
range of the inter-species scattering length. The direct quantitative comparison
with our work, however, is less straightforward because Law et al.'s analysis
assumes a spherically symmetric condensate, while we analyse a quasi one-dimensional
model.

\subsection{Static bright-bright soliton pairs\label{sec:StaticBBSolitonPairs}}

Let us now investigate the temporal stability of a static ($v=0$) bright-bright
soliton pair obtained as the exact solution of Eqs.~(\ref{eq:coupled_GP1}-b) in
the absence of a trapping potential ($\lambda_{i}=0$). The solutions have the
form as in Eq.~(\ref{sol-excit}) where $\varphi_{i}(x)$ are chosen to have the
functional forms given in Eqs.~(\ref{eq:_bb1}-b).

The existence conditions (\ref{eq:_BBcond}) prescribe the relations between
domains of the inter- and intra-species coupling strengths, $b_{ij}$, which
are listed in Table \ref{tab:ExistenceOfBBSolitons}. Moreover, the common
wave-vector is $k = 1/ (2 C_{1})$.

\begin{table}[h]
   \begin{tabular}{r<{\hspace*{1mm}}|p{5mm}p{5mm}cc}
      Case & $b_{11}$ & $b_{22}$ & Constraint on $b_{12}$ \\
      \hline\hline
      1    & $+$ & $-$ & $b_{12}< b_{22}/\kappa$ \\
      2    & $-$ & $+$ & $b_{12}< b_{11}/\kappa$ \\
      3a   & $-$ & $-$ & $b_{22}/\kappa < b_{12}<\sqrt{b_{11}b_{22}}$ & if $\kappa^2 b_{11}<b_{22} $\\
      3b   &      &      & $b_{12}<b_{11}\kappa$ & if $\kappa^2 b_{11}<b_{22} $ \\
      3c   &      &      & $b_{11}\kappa < b_{12}<\sqrt{b_{11}b_{22}}$& if $\kappa^2 b_{11}>b_{22} $\\
      3d   &      &      & $b_{12}<b_{22}/\kappa$ & if $\kappa^2 b_{11}>b_{22} $ \\
      4    & $+$ & $+$ & $b_{12}<-\sqrt{b_{11}b_{22}}$\\
   \end{tabular}
   \caption{\label{tab:ExistenceOfBBSolitons}
            Various domains of inter- and intra-atomic interaction strengths,
            $b_{ij}$, permitting the existence of a bright-bright soliton pair.
           }
\end{table}

\begin{figure}[htb]
   \includegraphics[width=\textwidth/2-3.7mm]{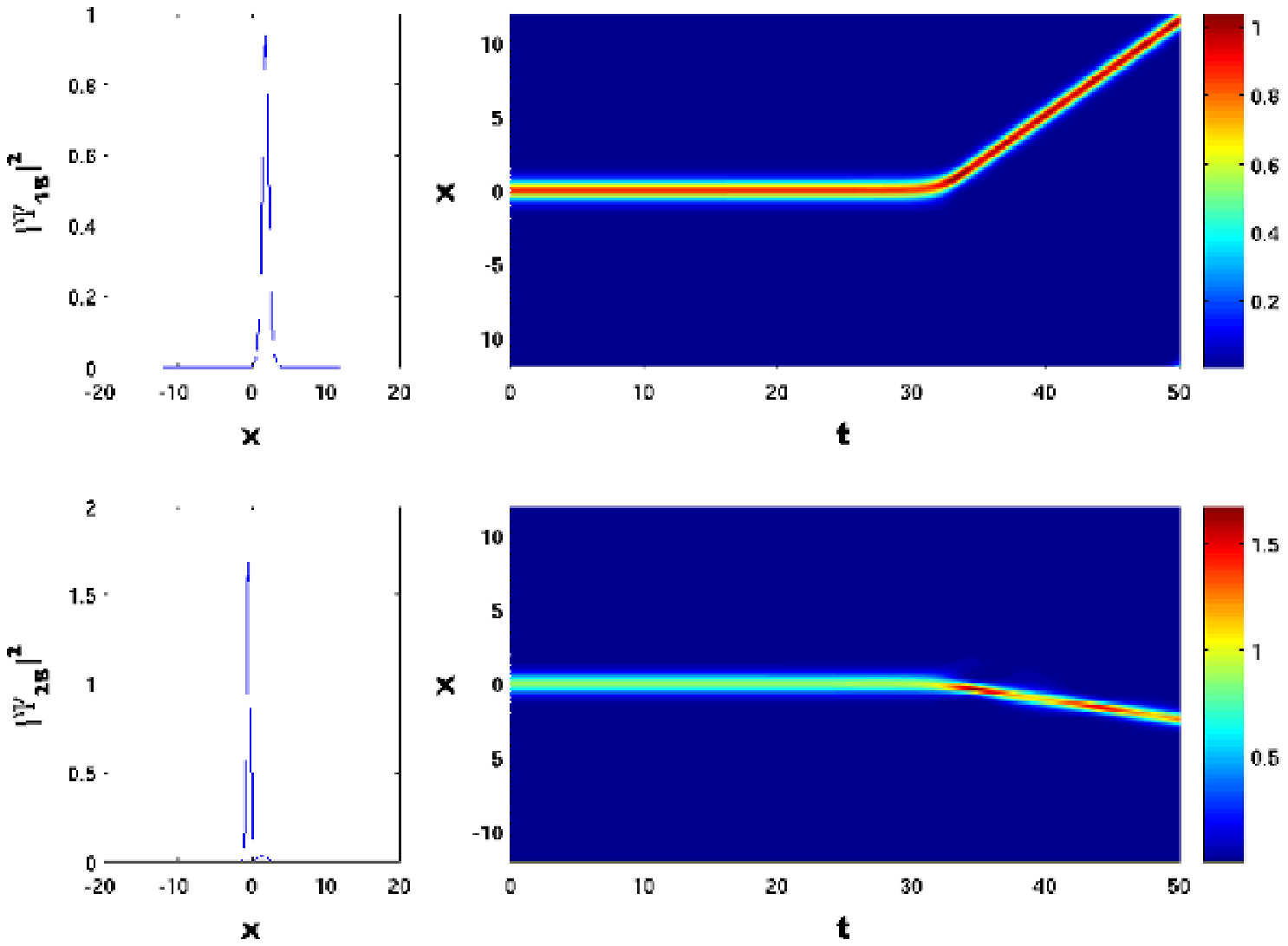}
   \includegraphics[width=\textwidth/2-3.7mm]{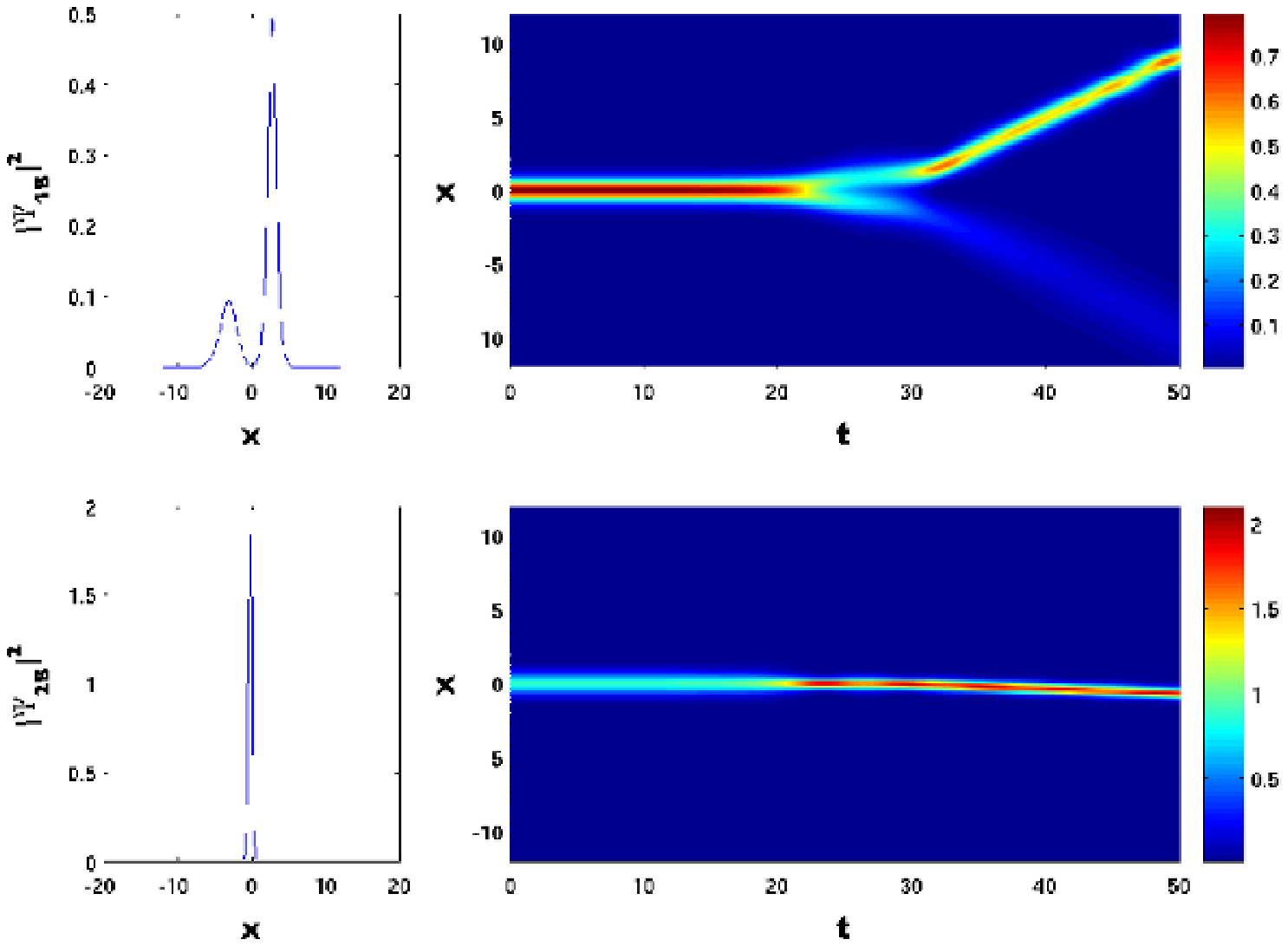}
   \includegraphics[width=\textwidth/2-3.7mm]{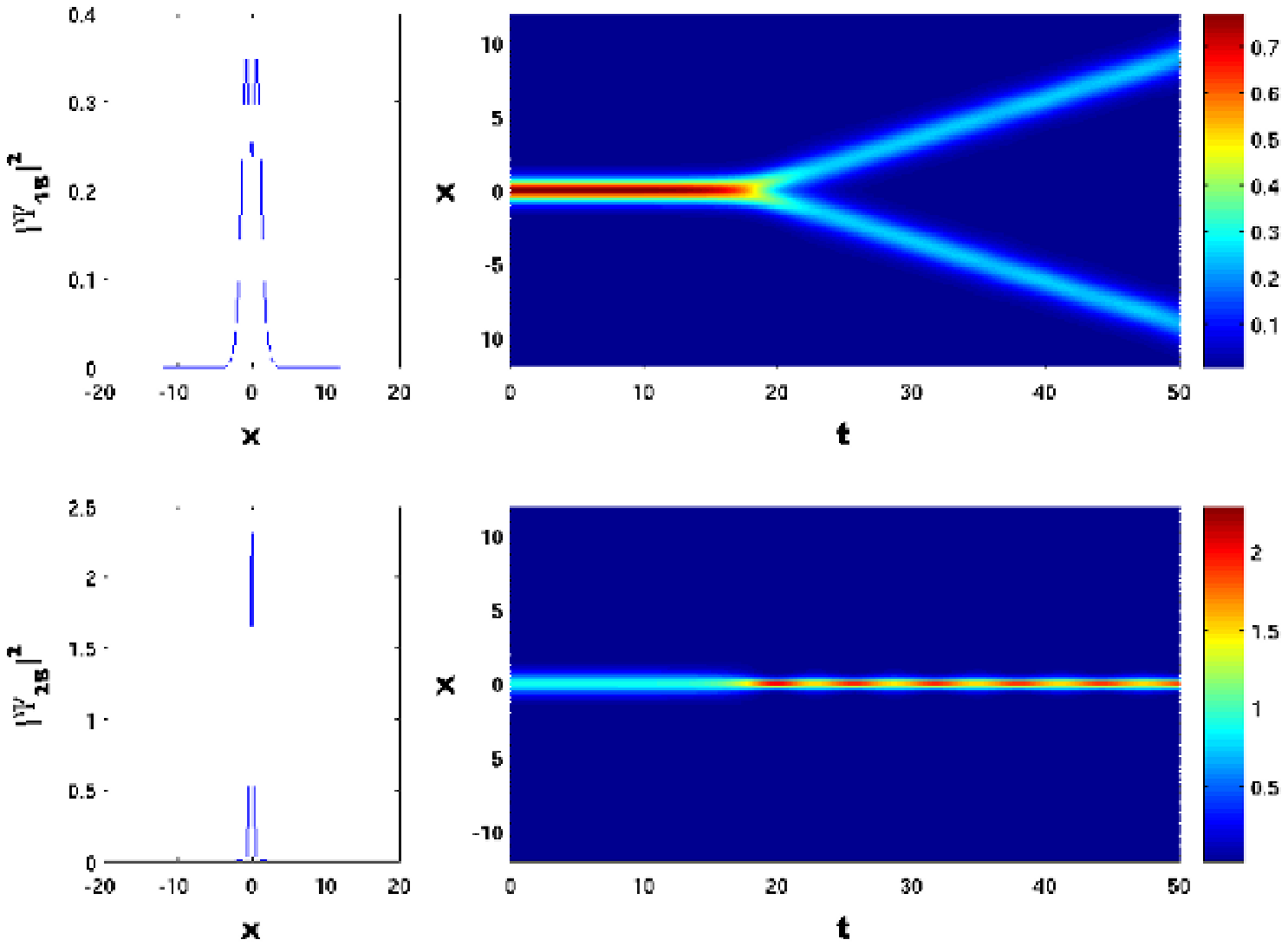}
   \caption{\label{fig:BrightBright}%
            (Color online) Evolution of a bright-bright soliton pair in a binary BEC
            composed of $^{7}$Li and $^{39}$K atoms. Intraspecies scattering lengths
            and atom numbers are $a_{12}=0.2$\,nm ($b_{12}=0.46$), $N_{2}=2029$ (first
            two panels), $a_{12}= 0.275$\,nm ($b_{12}=0.64$) $N_{2}=2270$ (second pair
            of the panels), $a_{12}= 0.3$\,nm ($b_{12}=0.7$) $N_2=2347$ (third pair of
            the panels). Other parameters are: $a_{11}=-1.4$\,nm $(b_{11}=-3.93)$,
            $a_{22}= -0.9$\,nm $(b_{22}=-1.07)$, $m_{1}=7$\,au, $m_{2}=39$\,au,
            $\omega_{1, \perp}=2\pi\times 710$\,Hz, $\lambda_{1} = \lambda_{2}=0$,
            $v=0$, $N_{1}=2000$. The snapshots are taken at $t=35$, $t=35$, and
            $t=20$, respectively.}
\end{figure}
\clearpage

Scenario 4, for example, describes two condensates for which the intra-species
interactions are repulsive. This situation, using Hartree-Fock calculation, has
been theoretically analysed \cite{Esry1997} soon after the observation of overlapping
condensates prepared from two hyperfine states of $^{87}$Rb \cite{Myatt1997}.
For this case, i.e. $m_{1}=m_{2}$, it was established that the two condensates
cannot co-exist if $\abs{a_{12}} < \sqrt{a_{11}a_{22}}$. Our approach reproduces
and extends this result for the case of different species. This surprisingly simple
relation can be understood using energetic arguments; if $b_{12}$ overcomes the
geometric mean of the intra-species interaction strengths, the repulsion between
the two condensates will separate the two condensate completely and they will
not overlap any more. However, if the two species attract each other enough,
i.e. $b_{12}< -\sqrt{b_{11}b_{22}}$, the attraction will dominate and can counteract
the individual repulsion present in each component.

Another interesting scenario here is the one listed under the case 3a in
Table \ref{tab:ExistenceOfBBSolitons} showing that it is possible to create
a bright-bright pair, within the range $0 < b_{12} < \sqrt{b_{11} b_{22}}$,
in spite of the repulsive inter-atomic interaction. In order to investigate
the stability of the bright-bright soliton pair in this domain we simulate
the temporal evolution of the BEC system composed of $^7$Li and $^{39}$K
atoms accessible for experiments. As Fig.~\ref{fig:BrightBright} shows,
depending on the value of the interspecies interaction, two types of
instability may occur. At values of $a_{12}$ less than a critical value of
about $0.275$\,nm, the two standing solitons begin to repel each other
(first two panels) and depart from each other as they were particles,
preserving the total zero momentum. Above the critical $a_{12}$ value
the soliton with constituents of the smaller mass splits into two equal
parts and the heavier component becomes a breather keeping its original
place (third pair of panels). At the critical value of $a_{12}$ both
types of instability can be observed (third and fourth panel)
simultaneously.

Similar instability effects of a bright-bright and other soliton pairs
have also been observed in Ref.~\cite{Beitia2010}. The instability is
explained by Kevrekidis at al.~\cite{Kevrekidis2005} as the appearance
of a negative eigenvalue pair obtained by a linear stability analysis.
It has also been shown that higher order bright-bright soliton pairs
would exhibit instability irrespectively of the system parameters.
Interestingly, it is possible for one of the higher-order bright solitons,
i.e., for the one which has the stronger self-attraction, to recover its
stability by collapsing into one bright soliton and expel the other
component from its original position.

Finally, it is important to note here, that the phases of the divided
solitons are equal, therefore these particle-like wave-packets remain
coherent with each other. Similarly to an optical beam splitter where
light is divided into two coherent beams, one could divide these matter
waves and use one of them as a probe and the other one as a control
packet. The probe packet could undergo transformations, while the control
packet is left to evolve freely, thereby a phase-difference could build
up between the two packets. If the two packets are brought together again,
the phase-difference could cause interference pattern which allows one to
quantify coherence. This may potentially be helpful for calibrational
purposes as well.

\subsection{Dark-dark soliton pairs \label{sec:DDSolitonPairs}}

\begin{figure}[hbt!]
   \includegraphics[width=\textwidth/2-3mm]{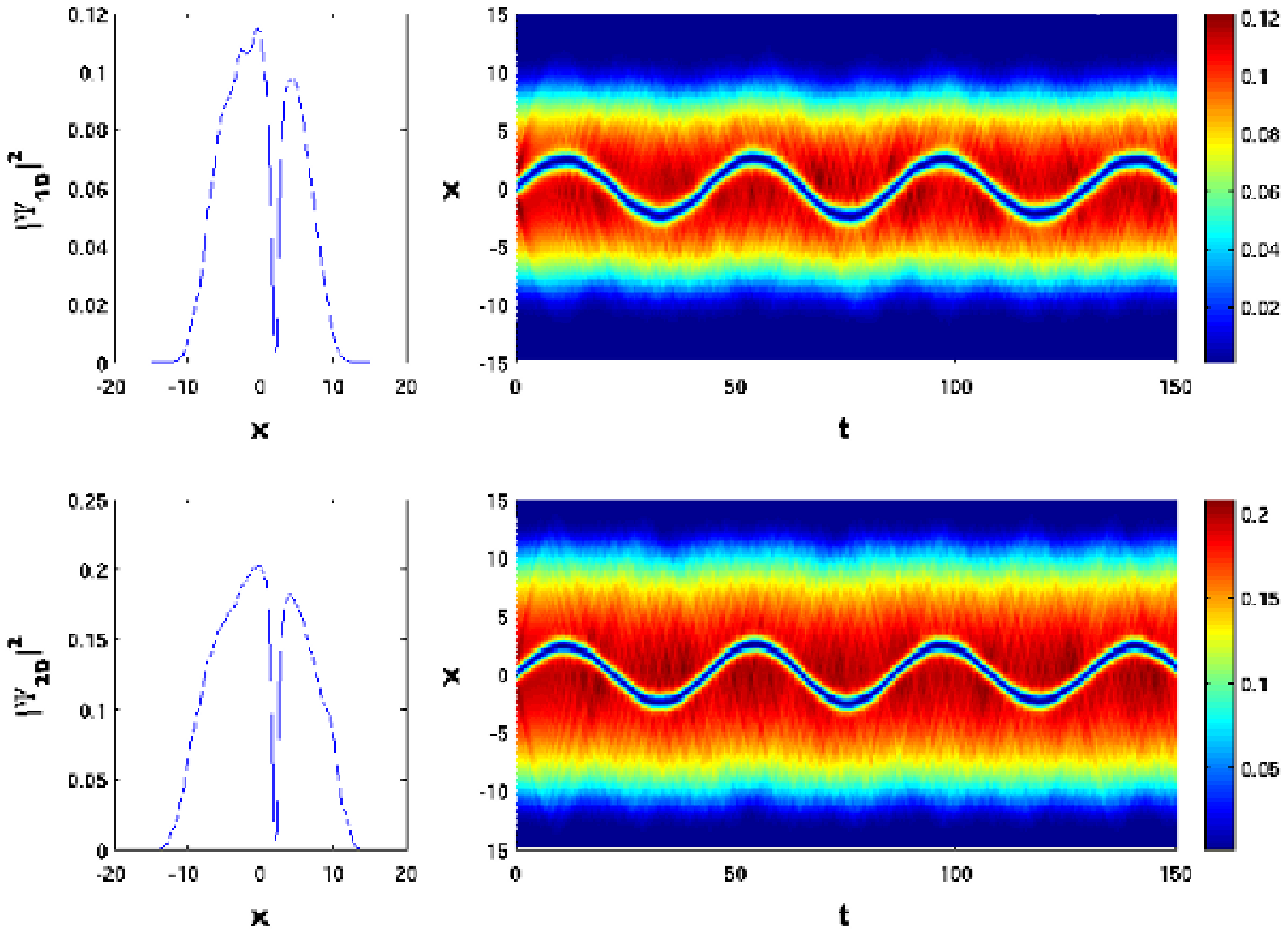}
   \includegraphics[width=\textwidth/2-3mm]{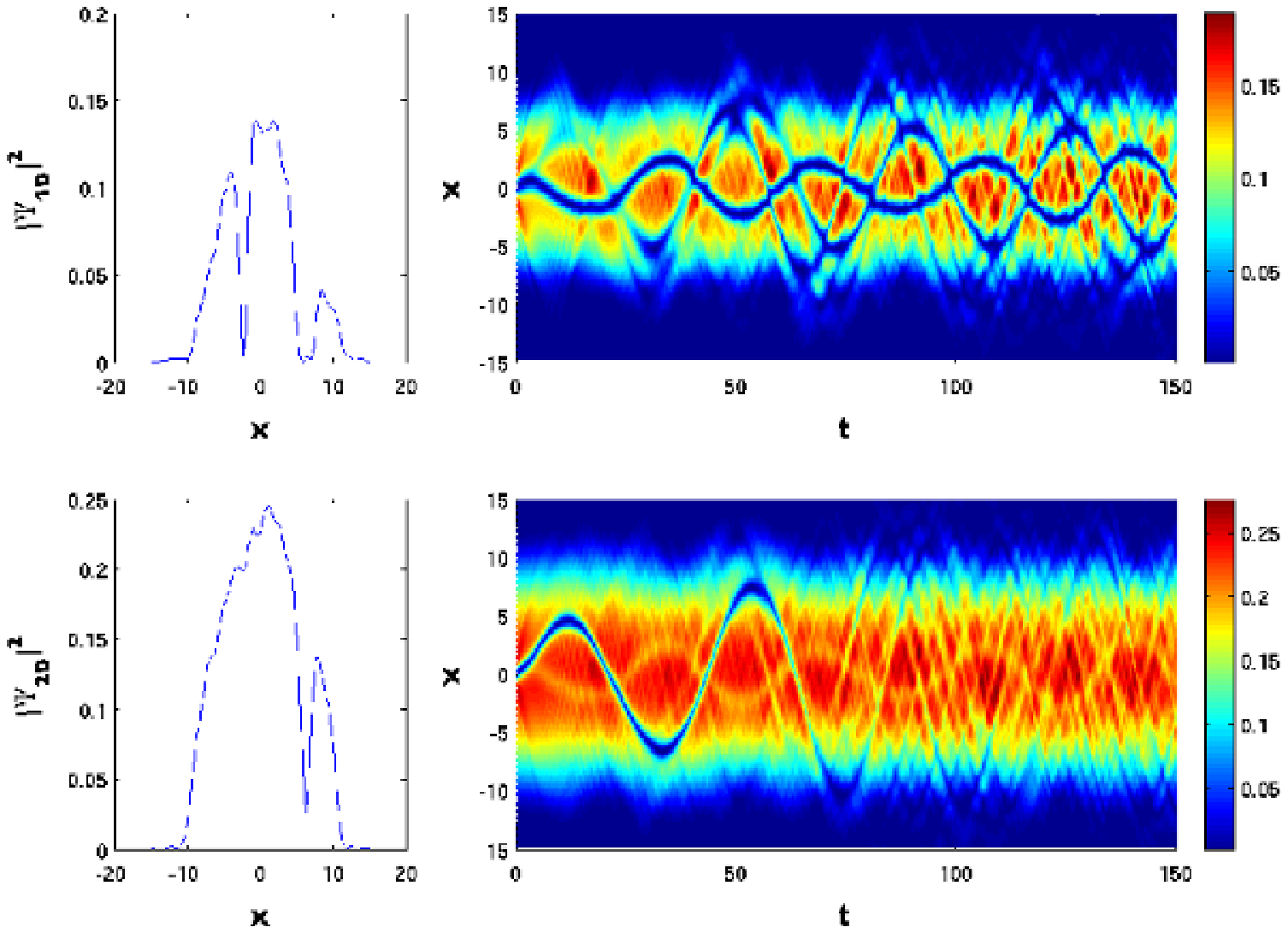}
   \caption{\label{fig:DarkDark}%
            (Color online) Evolution of a dark-dark soliton pair in a
            binary BEC composed of $^{87}$Rb and $^{87}$Rb atoms.
            Intraspecies scattering lengths are $a_{12}=1.5$\,nm
            ($b_{12}=3.62$) (top panel), $a_{12}= -1.5$\,nm ($b_{12}=
            -3.62$) (bottom panel). Other parameters are: $a_{11}=0.97
            \times 5.5$\,nm $(b_{11}=13.04)$, $a_{22}=1.03\times 5.5$\,nm
            $(b_{22}=13.52)$, $m_{1}=87$\,au, $m_{2}=87$\,au,
            $\omega_{1, \perp}=2\pi\times 710$\,Hz, $\lambda_{1} =
            \lambda_{2}=0.2$, $v=0.2$, $N_{1}=500$, $N_{2}=1600$.
            The snapshots are taken at $t=50$.
           }
\end{figure}

At the end we are examining the third possible combination of soliton pairs;
a dark soliton is excited in each condensate. Interestingly this pairing can
be stable even if the interaction inside each condensate is attractive ($a_{11}$,
$a_{22} <0$). The dark-dark solitons are described by the following formulas
\begin{widetext}
\begin{subequations}
   \begin{eqnarray}
      \widetilde{\psi}_{1}^{\mathrm{DD}}
      &=&
      \left \lbrack
         i \sqrt{-C_{1}} \,v +
         \Phi^{\mathrm{TF}}_{1} (0) \varphi_{1}^{\mathrm{DD}} (x-vt)
      \right \rbrack \,
      \exp{ \left ( -i {\widetilde{E}}_{1}^{\mathrm{DD}} t \right ) }
      \exp{ \left \lbrack -i \left( \frac{b_{12} C_{2}}{\kappa^2} - 
                          b_{11} C_{1} \right ) v^{2} t
            \right \rbrack},
      \\
      \widetilde{\psi}_{2}^{\mathrm{DD}}
      &=&
      \left [ i \frac{\sqrt{C_2}}{\kappa}\,v +
      \Phi^{\mathrm{TF}}_{2} (0)
      \varphi_{2}^{\mathrm{DD}}(x-vt)\right]\,
      \exp{\left(-i {\widetilde{E}}_{2}^{\mathrm{DD}} t\right)}
      \exp{\left[-i\left( \frac{b_{22} C_{2}}{\kappa^2}-b_{12} C_{1}
                   \right ) v^{2} t \right]}.
\end{eqnarray}
\end{subequations}
\end{widetext}
For our numerical investigations a $^{87}$Rb--$^{87}$Rb system has been
chosen, with repulsive intra-species interactions, $a_{11}=5.335$\,nm
and $a_{22} = 5.665$\,nm taken from \cite{Hall1998}. This choice of
coupling corresponds to scenario 4 of Table~\ref{tab:ExistenceOfDDSolitons}.
Tuning $a_{12}$ into the positive regime results in a stable dark-dark
soliton pairs, see top two panels of Fig.~\ref{fig:DarkDark}. In the
bright-bright case we have found that if the intra-species interactions
are attractive and the inter-species interactions are repulsive, the
solutions are unstable. We have tested numerically that this observation
is valid in the dark-dark case as well, if the sign of the interactions
is inverted. Furthermore, our numerical calculation provided an interesting
phenomenon in the dynamics of these solitons. Preparing the dark solitons
in the same way as before (i.e., by superimposing onto the ground state)
but reversing the sign of $a_{12}$, resulted not only in losing the
long-term stability, but also in the decay of a dark soliton in one
of the components, and emerging as a secondary dark soliton in the
other component (bottom two panels of Fig.~\ref{fig:DarkDark}).

\begin{table}[h]
   \begin{tabular}{r<{\hspace*{1mm}}|p{5mm}p{5mm}cc}
      Case & $b_{11}$ & $b_{22}$ & Constraint on $b_{12}$ \\
      \hline\hline
      1    & $+$ & $-$ &                        $b_{12} > b_{11} \kappa$        \\
      2    & $-$ & $+$ &                        $b_{12} > b_{22}/\kappa$        \\
      3    & $-$ & $-$ &                        $b_{12} > \sqrt{b_{11}b_{22}} $ \\
      4a   & $+$ & $+$ & $-\sqrt{b_{11}b_{22}} < b_{12} < b_{22}/\kappa$ & if $\kappa^2 b_{11}>b_{22}$\\
      4b   &     &     &                        $b_{12} > b_{11} \kappa$ & if $\kappa^2 b_{11}>b_{22}$\\
      4c   &     &     & $-\sqrt{b_{11}b_{22}} < b_{12} < b_{11}\kappa$  & if $\kappa^2 b_{11}<b_{22}$\\
      4d   &     &     &                        $b_{12} > b_{22}/\kappa$ & if $\kappa^2 b_{11}<b_{22}$\\
\end{tabular}
   \caption{\label{tab:ExistenceOfDDSolitons}
            Various domains of inter- and intra-atomic interaction strengths,
            $b_{ij}$, permitting the existence of a dark-dark soliton pair.
           }
\end{table}

\section{Summary \label{sec:summary}}

We have considered the existence and stability of soliton excitations in a
two-component Bose-Einstein condensate both analytically and numerically.
Our model allows the components to represent different elements ($m_{1}
\ne m_{2}$), but we also included those cases when two hyperfine states of
the same element constitute the condensates. We excluded the possibility
that these components can transmute into each other, i.e. the hyperfine
states cannot be driven into each other. The dynamics of these condensates,
within the mean-field zero-temperature approximation, are governed by the
coupled Gross-Pitaevskii equations. We chose our presentation to be suitable
for combining analytical results with earlier numerical investigations,
e.g. \cite{Schumayer2001, Xun-Xu2010, Liu2009}.

The occurrence of particle-like excitations together with conserved quantities
are associated with the integrability of a nonlinear evolution field-equation,
such as the coupled Gross-Pitaevskii equations. Using the results of a recent
Painlev{\'e} analysis of the coupled Gross-Pitaevskii equation
\cite{Schumayer2001}, we showed how the system parameters determine the
integrability of this system. However, for the CGP equations the studies
so far restricted themselves for either equal coupling coefficients ($b_{11}=
b_{12}=b_{22}$) and/or equal masses. We note here that excitations with
long lifetime may exist for a nonlinear evolution equation even when the
integrability conditions are violated \cite{Novoa2008}, but these cases
are possibly exceptional and do not represent the generic behaviour.

\begin{figure}[hbt!]
   \includegraphics[width=\textwidth/2-4mm]{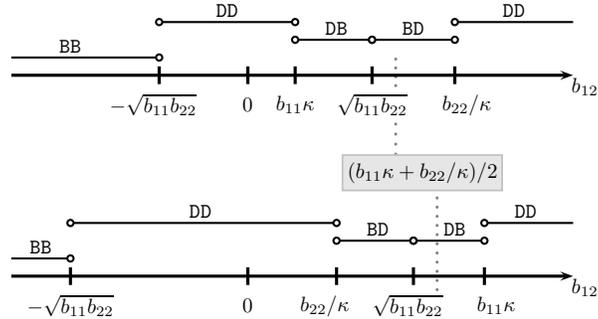}
   \caption{\label{fig:DifferentRepresentation}%
            Ranges of the inter-species coupling coefficient, $b_{12}$, in
            which different types of soliton pairs can exist and are stable.
            The top graph is valid if $\kappa^{2} b_{11} < b_{22}$, while
            the lower graph is for the complementary case, i.e. $\kappa^{2}
            b_{11} > b_{22}$. The dotted line shows the result of the
            Painlev{\'e}-test (\ref{eq:ConditionForIntegrability}) in
            the case of harmonic trapping potentials.
           }
\end{figure}

We examined those ranges of system parameters which permit coupled soliton
solutions of the governing equations (\ref{eq:coupled_GP1}-b). Here, we
utilised the two well-known one-soliton solutions of the one-component
nonlinear Schr{\"o}dinger equation; the bright and dark solitons. These
excitations then spatially modify the Thomas-Fermi ground state in our
theoretical description or by the appropriate ground state in our numerical
simulation. We found analytically that each possible pair of these
solitons have a range of system parameters where they are stable, and
presented these ranges in tabulated form in Tabs.
\ref{tab:ExistenceOfBDSolitons}, \ref{tab:ExistenceOfBBSolitons},
and \ref{tab:ExistenceOfDDSolitons}. One can present these
findings differently, if the intra-species coupling coefficients
are held relatively fixed, and only the inter-species one is
tuned. For example, if we assume $b_{11}$ and $b_{22}$ to be
positive (effective repulsive interaction) then the three tables
\ref{tab:ExistenceOfBDSolitons}, \ref{tab:ExistenceOfBBSolitons},
and \ref{tab:ExistenceOfDDSolitons} can be combined into one
graph (see Fig.~\ref{fig:DifferentRepresentation}).

Moreover, we also examined how the stability of these soliton pairs is
changing if the inter-species coupling coefficient is detuned from the
value predicted by the Painle{\'e}-analysis via
Eq.~(\ref{eq:ConditionForIntegrability}). It was shown, irrespectively
of the pairing, that the stability is lost, although the pairs were not
equally sensitive to the detuning, e.g. the motion of the bright-dark
pair became erratic and preserved its periodicity only qualitatively
after changing $a_{12}$ from 5.5\,nm to 5.4\,nm.

Two types of instability of static bright-bright soliton pairs composed
of species with unequal masses have been observed. When the interspecies
interaction lies below a critical value ($0 < b_{12} < b_{12}^{\mathrm{cr}}$),
the static bright-bright pair evolves into a repulsive, momentum conserving,
moving soliton pair. When the value of $b_{12} > b_{12}^{cr}$ then one of
the bright soliton (the constituent with smaller mass) splits into two equal
portion of same phase while the other bright soliton becomes a breather.

Well below the critical temperature of the Bose-Einstein condensate our
description is expected to be adequate and the results could help
experimentalist to modify the scattering lengths via Feshbach resonance
into a range where stable soliton pairs exist. As the temperature increases,
however, the interaction with the thermal cloud becomes more and more
important, and could not be neglected any more. We have not yet examined
how the interplay between the condensates and the thermal clouds (for each
component) would modify the dynamics. This needs further research beyond
the mean field description.

{\emph{Acknowledgment}} ---  D. Schumayer acknowledges financial
support from NERF-UOOX0703 (NZ) and also by the University of
Otago. G. Csire and B. Apagyi thank DFG for subsidising their
stay at Institute of Theoretical Physics, University of Giessen.


%

\end{document}